\documentclass[%
preprint,
 amsmath,amssymb,
 aps,
]{revtex4-1}

\usepackage{graphicx}
\usepackage{dcolumn}
\usepackage{bm}

\newcommand{\vect}[1]{\mbox{\boldmath${#1} $}} 
\newcommand{\vex}{{\vect x}}

\newcommand{\la}{\langle}
\newcommand{\ra}{\rangle}
\newcommand{\laa}{\langle\langle}
\newcommand{\raa}{\rangle\rangle}

\begin{document}


\title{On the Coefficients of \\a Hyperbolic Hydrodynamic Model}

\author{Shin Muroya}
\email{muroya@matsu.ac.jp}
\affiliation{%
Department of Comprehensive Management, 
Matsumoto University  \\
Matsumoto 390-1295, Japan
}

\author{Masashi Mizutani}
\email{mizutani@aoni.waseda.jp}
\affiliation{
Department of physics and Research Institute for Science and Engineering,\\
 Waseda University \\
 Tokyo 169-8555, Japan
}%


\date{\today}

\begin{abstract}
Based on the Nakajima-Zubarev type nonequilibrium density operator, we derive a hyperbolic hydrodynamical equation.  Microscopic Kubo-formulas for all coefficients in the hyperbolic hydrodynamics are obtained. Coefficients $\alpha_{i}$'s and $\beta_{i}$'s in the Israel-Stewart equation are given as current-weighted correlation lengths which are to be calculated in statistical mechanics.  
\begin{description}
\item[PACS numbers]
05.60.-k, 
12.38.Mh 
24.10.Nz 
\end{description}
\end{abstract}

\maketitle



\section{Introduction}

The hydrodynamic model is one of the widely applied phenomenological models for  relativistic heavy ion collisions.  
According to the recent detailed analyses, 
weak but non-vanishing viscosity is significant for
the qualitative discussion of $v_2$ at RHIC \cite{Hirano:2008hy}.
Because the Navier-Stokes equation is a parabolic type equation, naive relativistic 
extension is not consistent with causality, and as a result, the 
numerical solution becomes unstable \cite{HL}. 
The relativistic causal hydrodynamic model which is a hyperbolic type had been introduced phenomenologically by Israel and Stewart more than thirty years ago \cite{IS1, IS2}.  In the past decades, many proposals 
 appear for 
the appropriate equations of the relativistic causal hydrodynamic model \cite{TK, KK, MH}.
Most of the works seem to belong to the semi-classical phenomenological approach 
based on the Boltzmann equation.  In this paper we derive a hyperbolic type
 hydrodynamics as a second order hydrodynamics based on a nonequilibrium density operator method \cite{ISMD}.
Romatschke and Moore et al.~ also discussed hydrodynamics based on a nonequilibrium entropy density \cite{Romatschke,Moore}. In the present paper, 
more explicit formulas for the coefficients are issued. 

The hydrodynamic model is a phenomenological model of the macroscopic point of view
which is based on the coarse graining. The total system is constructed as a patchwork of a local system corresponding the statistical systems in equilibrium.  
The local system at each space-time point is expected to be microscopically 
large but macroscopically one point and moving with four velocity $U^{\mu}$.  
Thermodynamical quantities and the coefficients in the hydrodynamical equation 
are the local quantities as the 
functions of position through the local thermodynamical parameters such as 
temperature $T(x)$ and chemical potential $\mu(x)$.  
The macroscopic model based on the coarse graining is 
justified by the fact that the microscopic correlation length
is much smaller than the typical scale of the macroscopic dynamics.  
More than fifty years ago, Iso et al. 
discussed the applicability condition of the hydrodynamic model for 
the multiple production in high energy pp collisions based on the comparison
between the length of 
the canonical correlation of the current operators and the scale of the change 
in the solution of the hydrodynamical equation \cite{IMN}.

Regarding the self-consistent macroscopic model, before we start to
solve the hydrodynamical equation, all functional forms of 
the coefficients should be fixed.
All coefficients of the hydrodynamical equation are given as functions of 
thermodynamical parameters, $T$ and $\mu$.  The derivation of them is 
a major task for statistical mechanics.
The well-known Kubo formulas for the viscosities and heat conductivity  
provide us  ways to evaluate the coefficients of the Navier-Stokes equation 
by statistical mechanics in equilibrium.  
However, the method to calculate the additional coefficients in
the hyperbolic hydrodynamical equation, the so called Israel-Stewart equation, have not yet been  established.  As far as the authors know, only the kinetic 
calculation by using the Grad 14 moments method is adopted.
The aim of this paper is to establish a way to calculate all coefficients in 
the hyperbolic hydrodynamics as the functions of temperature and chemical potential. 
Based on the nonequilibrium density operator method, we systematically derive microscopic formulas for the coefficients $\alpha_i$'s and $\beta_i$'s in the Israel-Stewart equation.  
All coefficients are expressed as the current-weighted length of canonical correlation
which are to be calculated in statistical mechanics in equilibrium with $T$ and $\mu$.  
In general, evaluation of the canonical correlation is not easy, but our formulas can be good targets for the future Lattice QCD simulations and the hadro-molecular calculation.

\section{hydrodynamic model}

Hydrodynamical equation is composed of the energy-momentum conservation law
\begin{equation}
\partial^{\mu} T_{\mu \nu} = 0
\end{equation}
and the charge conservation law 
\begin{equation}
\partial^{\mu} J_{\mu} = 0.
\end{equation}
The flow of the fluid is described by a four velocity $U^{\mu}$ which is a normalized time-like vector,
$U^{\mu}U_{\mu} = 1$.  By using $U^{\mu}$, the space-like projection operator is defined as 
$\Delta^{\mu \nu} = g^{\mu \nu}- U^{\mu}U^{\nu}$. 
With $U^{\mu}$ and $\Delta^{\mu \nu} $, we can define four projection operators 
of the second rank symmetric tensor:
\begin{eqnarray}
\mathcal{T}(^{\mu\nu}|^{\rho\sigma})&=& \frac{1}{2}\left(\Delta^{\mu\rho}\Delta^{\nu\sigma}+\Delta^{\mu\sigma}\Delta^{\nu\rho}
-\frac{2}{3}\Delta^{\mu\nu}\Delta^{\rho\sigma}\right),
\\
\mathcal{S}(^{\mu\nu}|^{\rho\sigma}) &=& \frac{2}{3}\Delta^{\mu\nu}\Delta^{\rho\sigma},
\\
\mathcal{V}(^{\mu\nu}|^{\rho\sigma}) &=& \frac{1}{2}\left( U^{\mu}U^{\rho}\Delta^{\nu\sigma}+\Delta^{\mu\rho}U^{\nu}U^{\sigma} 
+ U^{\mu}U^{\sigma}\Delta^{\nu\rho}+\Delta^{\mu\sigma}U^{\nu}U^{\rho} 
\right),
\\
\mathcal{U}(^{\mu\nu}|^{\rho\sigma}) &=& U^{\mu}U^{\nu}U^{\rho}U^{\sigma}.
\end{eqnarray}
$\mathcal{T}(^{\mu\nu}|^{\rho\sigma})$ stands for the traceless part of the spatial component,
$\mathcal{S}(^{\mu\nu}|^{\rho\sigma}) $ stands for the trace of the spatial component,
$\mathcal{U}(^{\mu\nu}|^{\rho\sigma}) $ stands for the time-like time-like component, and
$\mathcal{V}(^{\mu\nu}|^{\rho\sigma}) $ stands for the time-like space-like 
or the space-like time-like component, respectively.
These operators form a complete set
\begin{equation}
\mathcal{S}(^{\mu\nu}|^{\rho\sigma}) +\mathcal{T}(^{\mu\nu}|^{\rho\sigma})+
\mathcal{U}(^{\mu\nu}|^{\rho\sigma}) +\mathcal{V}(^{\mu\nu}|^{\rho\sigma}) =
\frac{1}{2} \left( g^{\mu\rho}g^{\nu\sigma}+g^{\mu\sigma}g^{\nu\rho} \right) = \openone .
\end{equation}

$U^{\mu}$ corresponds to the direction of the time on the local rest frame at $x^{\mu}$ and $\Delta^{\mu\nu}$ specifies the space-like coordinate, respectively.  Hence, on the local rest frame,  
 $U^{\mu}$ becomes $(1,0,0,0)$ and $\Delta^{\mu\nu}$ becomes $-\delta^{ij} $, respectively.  
 Throughout this paper, Greek indexes, $ \mu, \nu,\rho, \sigma$ etc.\ stand for the Lorentzian (0,1,2,3), and  Latin indexes $ i,j,k$ etc.\ stand for the Euclidian (1,2,3), respectively.
A space-time point $(\vex,t)$ is denoted as $x^{\mu}$ for the abbreviation.  $\tau =  x^{\mu}U_{\mu}$ 
stands for the time on the local rest frame (comoving frame of the fluid element), and 
$D = U^{\mu} \partial_{\mu}$ stands for the derivative by $\tau$, 
respectively.

Energy density is given as the time-like time-like component of the energy-momentum tensor and pressure is given as the the trace of the spatial part,  respectively,
\begin{eqnarray}
\varepsilon &=& T^{\mu\nu} U_{\mu}U_{\nu},  \\
 P &=&\frac{1}{3}T^{\mu\nu}\Delta_{\mu\nu}.
\end{eqnarray}

Once Lagrangian of the system is given, the energy-momentum tensor $T^{\mu\nu}$ 
is defined as the space-time translation operator.  
However, the definition of the flow 
$U^{\mu}$ of a relativistic fluid is not trivial.  
Eckart \cite{Ekt} and Namiki et al. \cite{NI} define $U^{\mu}$ based on the particle flow (P-frame) and Weinberg used charge current \cite{Wb}.  In the textbook by Landau and Lifshitz, $U^{\mu} $ is defined as a eigenvector of $T^{\mu\nu}$ \cite{LL}.  Recently another possible choice is proposed by Tsumura et al.\ \cite{TKO}. 
In this paper we adopt the Landau-Lifshitz frame (E-frame) where $U^{\mu}$ is an eigenvector of the energy momentum tensor:
\begin{equation}
T^{\mu \nu }U_{\nu}. =  \varepsilon  U^{\mu}. 
\label{E-frame}
\end{equation}
Stress-shear tensor is given as the spatial traceless part of the energy momentum tensor:
\begin{equation}
\pi^{\mu \nu} =\mathcal{T}(^{\mu\nu}|_{\rho\sigma}) T^{\rho \sigma}.
\end{equation}
Using the projection operators, we can denote energy density as:
\begin{equation}
\varepsilon =  U_{\mu}U_{\nu}   \mathcal{U}(^{\mu\nu}|_{\rho\sigma}) T^{\rho\sigma},
\end{equation}
and pressure as:
\begin{equation}
P = \frac{1}{3} \Delta_{\mu\nu} \mathcal{S}(^{\mu\nu}|_{\rho\sigma}) T^{\rho\sigma},
\end{equation}
respectively.
The time-like space-like components of the energy momentum tensor 
should correspond to the heat flow, however, because of Eq.\ (\ref{E-frame})
these components must vanish:
\begin{equation}
\mathcal{V}(^{\mu\nu}|^{\rho\sigma}) T^{\rho\sigma} = 0,
\end{equation}
on the E-frame.

The charge density $n$ and the charge current $J^{\mu}$ are given by 
$n= J^{\mu}U_{\mu}$ as usual, however,  
$U^{\mu} $ is proportional to the energy flow on the E-frame, and 
the charge current contains $I^{\mu}$ which is perpendicular to $U^{\mu}$:
\begin{eqnarray}
J^{\mu} &=&  n U^{\mu} + I^{\mu}, \\
I^{\mu} &=&\Delta^{\mu}_{\nu}J^{\nu}.
\end{eqnarray}
$I^{\mu}$ corresponds to the {\it heat flow on the E-frame} \cite{LL}.

\section{Nonequilibrium density operator}
 
The nonequilibrium density operator is given as \cite{KYN,Zuba,CH}:
\begin{equation}
 \hat{\rho}= Q^{-1} {\rm exp}( -\hat{A} + \hat{B}) ,
\end{equation}
with $\hat{A}$ being:
\begin{equation}
\hat{A} =\displaystyle{\int} dS^{\mu}
\left\{
\beta(\vex,t) U^{\nu}(\vex,t)\hat{T}_{\mu \nu}(\vex,t)-
\beta(\vex,t) \mu(\vex,t) \hat{J}_{\mu}(\vex,t) \right\},
\end{equation}
$\hat{B}$ being:
\begin{eqnarray}
\hat{B} & = &   \lim _{\zeta \to 0+}\displaystyle{\int} d^{4} {x'}
{\rm e}^{-\zeta U^{\mu}(x-x')_{\mu}}
\left\{ \hat{T}_{\mu \nu}(\vex,'t')\partial'^{\mu}(\beta(\vex',t') U^{\nu}(\vex',t')) \right.  \nonumber \\
 &&      
\left.+ \hat{J}_{\nu}(\vex',t')\partial'^{\nu}(\beta(\vex',t') \mu(\vex',t')) \right\} \Bigg], \quad
U^{\mu}(x-x')_{\mu} \ge 0
\end{eqnarray}
and with $Q$ being normalization $Q = {\rm tr} [ {\rm exp}( -\hat{A} + \hat{B})  ]$.  The expectation value of an operator $\hat{O}(x)$ is given by $\langle \hat{O}(x) \rangle = {\rm tr}\left[ \hat{\rho} \hat{O}(x) \right]$. 

$\hat{A}$ corresponds to the local equilibrium density operator,
$\hat{\rho}_l=Q^{-1}_{l} {\rm exp} (-\hat{A}), $
with $ Q_{l}^{-1} = {\rm tr}\left[  {\rm exp} (-\hat{A}) \right] $ being normalization. 
$dS^{\mu}$ is a three-dimensional hypersurface which locally 
corresponds to the comoving volume element of the fluid.
The so-called matching conditions,
$ \la \hat{\varepsilon} \ra = \la \hat{\varepsilon}\ra_{l} $ and 
 $ \la \hat{n} \ra = \la \hat{n}\ra_{l} $,
are imposed for $\beta(\vex,t)$ and $\beta(\vex,t) \mu(\vex,t)$ in $\hat{\rho}_l$.  
If the expectation value of some operator $\hat{O}(x^{\mu})$:
\begin{equation}
\la \hat{O}(x^{\mu}) \ra_{l} = Q_{l}^{-1} {\rm tr} [\hat{\rho}_{l} \hat{O}(x^{\mu})],
\end{equation}
is a local quantity in the macroscopic sense, on the local rest frame at that point,  
the expectation value is the same to the one in the corresponding equilibrium state where 
the inverse temperature is $1/T =\beta(\vex,t) $ and the chemical potential is $\mu(\vex,t)$, 
respectively. 

$\hat{B}$ stands for the influence of the thermodynamical forces such as the disturbances in the flow and the thermodynamical parameters. 
The adiabatic limit $\zeta \to 0+$  should be taken at the final stage of the calculation.

According to the standard procedure of the linear response theory, we 
expand the nonequilibrium density operator $\hat{\rho}$ at around the local equilibrium 
operator $\hat{\rho}_l$  up to the linear term of the thermodynamical forces.  
Expectation value $\langle \hat{O}(x) \rangle$  is given as a canonical 
correlation with the force term:
\begin{eqnarray}
\langle \hat{O}(x) \rangle =\langle \hat{O}(x) \rangle_{l}  
&+& \displaystyle{
\lim _{\zeta \to 0+}\displaystyle{\int} d^{4} {x'}
{\rm e}^{-\zeta U^{\mu}(x-x')_{\mu}} }
\left(\hat{O}(x),\hat{T}_{\rho \sigma}(x'){\partial'} ^{\rho}\left(\beta(x')U^{\sigma}(x')\right) \right) \nonumber \\
&+& \displaystyle{
\lim _{\zeta \to 0+}\displaystyle{\int} d^{4} {x'}
{\rm e}^{-\zeta U^{\mu}(x-x')_{\mu}} }
\left(\hat{O}(x),\hat{J}_{\rho }(x'){\partial' }^{\rho}\left(\beta(x')\mu(x')\right) \right), 
\label{integration}
\end{eqnarray}
with $(\hat{O}_1,\hat{O}_2)$ being: 
\begin{equation}
(\hat{O}_1,\hat{O}_2)=\langle \hat{O}_1 \displaystyle{\int_{0} ^{1}}d\lambda e^{\lambda A} \hat{O}_2 e^{-\lambda A} \rangle_{l}
- \langle \hat{O}_1 \rangle_{l} \langle \hat{O}_2 \rangle_{l}.\label{CC}
\end{equation}

\section{Derivative expansion with the thermodynamical forces}

If the change of the thermodynamical parameters and the flow 
are slow enough and can be treated as 
almost constant during the microscopic relaxation length  
which is determined by the canonical correlation Eq.\ (\ref{CC}), we may
adopt the Taylor expansion of the thermodynamical forces  
${\partial'}^{\rho}\left(\beta(x')U^{\sigma}(x') \right)$  and 
${\partial'}^{\rho}\left(\beta(x')\mu(x')\right)$
at around the position $x$ of the operator $\hat{O}(x)$, and 
put the forces outside of the integration in Eq.\ (\ref{integration}).  
Then we can obtain expectation values of $T^{\mu\nu}$ and $J^{\mu}$ in the 
power series of the
gradient of the  thermodynamical parameters and the flow:
\begin{eqnarray}
\langle \hat{T}^{\mu \nu}(x) \rangle &=&\langle \hat{T}^{\mu \nu}(x) \rangle_{l} 
 + \left( \int d^3 \vex ' \int^{t}_{-\infty} dt' {\rm e} ^{-\zeta(t-t')}  (\hat{T}^{\mu \nu}(x),
\hat{T}_{\rho \sigma}(x')) \right) \partial ^{\rho}  \left(\beta U^{\sigma}\right)  \nonumber \\
 &&+ \left(\int d^3 \vex '\int^{t}_{-\infty} dt' {\rm e} ^{-\zeta(t-t')}  (\hat{T}^{\mu \nu}(x),
\hat{J}_{\rho}(x'))  \right)
\partial ^{\rho} \left(\beta \mu\right) \nonumber \\
&&+\left( \int d^3 \vex ' \int^{t}_{-\infty} dt' {\rm e} ^{-\zeta(t-t')} (\hat{T}^{\mu \nu}(x),({x_{\lambda}}'-x_{\lambda})
\hat{T}_{\rho \sigma}(x')) \right)
\partial ^{\lambda} \partial ^{\rho}  \left(\beta U^{\sigma}\right) \nonumber \\
&&+ \left( \int d^3 \vex ' \int^{t}_{-\infty} dt' {\rm e} ^{-\zeta(t-t')} (\hat{T}^{\mu \nu}(x),({x_{\lambda}}'-x_{\lambda})
\hat{J}_{\rho}(x'))\right) \partial ^{\lambda}
\partial ^{\rho}\left(\beta \mu \right), \label{tmn}
\label{expand1}
\end{eqnarray}
\begin{eqnarray}
\langle \hat{J}^{\mu}(x) \rangle &=& \langle \hat{J}^{\mu}(x) \rangle_{l}  
+ \left( \int d^3 \vex ' \int^{t}_{-\infty} dt' {\rm e} ^{-\zeta(t-t')}  (\hat{J}^{\mu}(x),
\hat{J}_{\rho}(x')) \right)
\partial ^{\rho} \left(\beta \mu\right)  \nonumber \\
&& + \left( \int d^3 \vex ' \int^{t}_{-\infty} dt' {\rm e} ^{-\zeta(t-t')}  (\hat{J}^{\mu}(x),
\hat{T}_{\rho \sigma}(x')) \right)
\partial ^{\rho} \left(\beta U^{\sigma}\right) \nonumber \\
&&+\left( \int d^3 \vex ' \int^{t}_{-\infty} dt' {\rm e} ^{-\zeta(t-t')} (\hat{J}^{\mu}(x),({x_{\lambda}}'-x_{\lambda})
\hat{J}_{\rho}(x')) \right) \partial ^{\lambda}
\partial ^{\rho}\left(\beta \mu \right) \nonumber \\
&&+\left( \int d^3 \vex ' \int^{t}_{-\infty} dt' {\rm e} ^{-\zeta(t-t')} (\hat{J}^{\mu}(x),({x_{\lambda}}'-x_{\lambda})
\hat{T}_{\rho \sigma}(x')) \right)
\partial ^{\lambda}
\partial ^{\rho}\left(\beta U^{\sigma}\right). \label{jm}
\label{expand2}
\end{eqnarray}
Here all coefficients are given as the integral of the canonical correlation 
with the local equilibrium density operator (\ref{CC}).  If the non-vanishing 
region of the canonical correlation in the integrand are limited only in 
the microscopic scale, we can regard them as local quantities in the 
macroscopic sense and we may substitute a statistical mechanical calculation in 
equiliblium with the corresponding $T$ and $\mu$ for the expectation 
with $\rho_{l}$ in Eqs.\ (\ref{tmn}) and (\ref{jm}).  

The first terms in the right hand side of (\ref{tmn}) and (\ref{jm}), which are free from the thermodynamical forces, 
correspond to the perfect fluid part: $
\langle \hat{T}^{\mu \nu}(x) \rangle_{l} = (\la \hat{\varepsilon} \ra_{l}+\la\hat{P}\ra_{l})U^{\mu}U^{\nu} - \la\hat{P}\ra_{l}g^{\mu \nu}$ 
and
 $\langle \hat{J}^{\mu}(x) \rangle_{l} = \la\hat{n}\ra_{l} U^{\mu}$.
The second terms in the right hand side of (\ref{tmn}) and (\ref{jm}), which are the 
the zeroth order gradient terms of thermodynamical forces, 
correspond to the Navier-Stokes equation.  The Kubo-formulas 
for viscosities and heat conductivity are obtained as 
the coefficients in these terms \cite{KYN,Zuba, NI}. 
The third terms in the right hand side of (\ref{tmn}) and (\ref{jm}), which are the 
the first order gradient terms of thermodynamical forces, 
correspond to the hyperbolic hydrodynamical equation \cite{ISMD}.

Decomposing $\hat{T}^{\mu\nu}$ into the space-like components and the time-like components, we can rewrite $\hat{B}$ as:
\begin{eqnarray}
\hat{B} = && \lim _{\zeta \to 0+}\displaystyle{\int} d^{4} {x'}
{\rm e}^{-\zeta U^{\mu}(x-x')_{\mu}}
\left\{ \beta(\vex',t') \hat{\pi}_{\mu \nu}(\vex,'t')\partial'^{\mu}( U^{\nu}(\vex',t')) \right.  \nonumber \\ \noalign{\vskip 0.4cm}
 &&      
- \hat{I}_{\nu}(\vex',t')\partial'^{\nu}(\beta(\vex',t') \mu(\vex',t'))  
\nonumber \\ \noalign{\vskip 0.4cm} 
 && 
- \left. \beta(\vex',t') \hat{P'}(\vex,'t') 
\partial_{\mu} U^{\mu}(\vex',t')  \right\},  
\end{eqnarray}
where:
\begin{equation}
\hat{P'} = \hat{P}-\left(\displaystyle{\frac{\partial \langle P \rangle_l}{\partial \langle \varepsilon \rangle _l}} \right)_{\langle n \rangle_l } 
\hat{\varepsilon} - \left(\displaystyle{\frac{\partial \langle P \rangle_l}{\partial \langle n \rangle_l} } \right)_{\langle \varepsilon \rangle_l } \hat{n}. \label{P'}
\end{equation}
By virtue  of the matching condition and the E-frame condition (\ref{E-frame}),   $\mathcal{S}(^{\mu\nu}|_{\rho\sigma})T^{\rho\sigma}$ and $\mathcal{V}(^{\mu\nu}|_{\rho\sigma})T^{\rho\sigma}$ don't appear in $\hat{B}$.

Let us denote integration of the canonical correlation in Eqs.\ (\ref{expand1}), (\ref{expand2}) as:
\begin{equation}
\laa O_{1} | O_{2} \raa = \lim _{\zeta \to 0+}
\int d^{3}\vex ' \int_{-\infty} ^{t}dt' {\rm e}^{-\zeta(t-t')} \left( \hat{O_1}(\vex,t),  \hat{O_2}(\vex',t') \right), 
\end{equation}
for the abbreviation. Assuming isotropy of the local rest system,  we can apply 
Curie's theorem for $\laa O_1 | O_2 \raa$.  $\laa O_{1} | O_{2} \raa $ depends on $ (\vex,t)$ only through the temperature and the chemical potential.   
Because of the translational invariance of 
the corresponding equilibrium system, we can rewrite:
\begin{eqnarray}
&&\lim _{\zeta \to 0+}
\int d^{3}\vex ' \int_{-\infty} ^{t}dt' {\rm e}^{-\zeta(t-t')} \left( \hat{O_1}(\vex,t),  (x'^{\mu}-x^{\mu})\hat{O_2}(\vex',t') \right), \\
&=& \lim _{\zeta \to 0+}
\int d^{3}\vex  \int_{-\infty} ^{0}dt {\rm e}^{+ \zeta t} \left( \hat{O_1}({\bf 0},0),  x^{\mu} \hat{O_2}(\vex,t) \right), \\
 \noalign{\vskip 0.4cm}
&=& \laa O_{1}|x^{\mu}|O_{2} \raa.
\end{eqnarray}
  
Substituting the projection operator $g^{\mu\nu} = U^{\mu}U^{\nu}+\Delta^{\mu\nu}$ into equations (\ref{expand1}) and (\ref{expand2}), and applying the Curie's theorem, we can obtain simple forms of the expectation values of the thermodynamical currents:
\begin{eqnarray}
\la\hat{\pi}^{\mu\nu}\ra 
&=& \la\hat{\pi}^{\mu\nu}\ra_{l} + \beta \laa \pi^{\mu\nu} | \pi^{\rho\sigma}\raa \partial_{\rho} U_{\sigma} - \laa \pi^{\mu\nu} | I^{\rho}\raa \partial_{\rho}(\beta \mu) \nonumber \\ 
&&+ \beta \laa \pi^{\mu\nu} |x^{\lambda} | \pi^{\rho\sigma}\raa \partial_{\lambda} \partial_{\rho} U_{\sigma} 
- \laa \pi^{\mu\nu} |x^{\lambda} | I^{\rho}\raa \partial_{\lambda} \partial_{\rho}(\beta \mu)  \nonumber \\
&=& 
 \la\hat{\pi}^{\mu\nu}\ra_{l} + \beta \laa \pi^{\mu\nu} | \pi^{\rho\sigma}\raa \partial_{\rho} U_{\sigma} 
\nonumber \\
&&
+ \beta \laa \pi^{\mu\nu} |\tau | \pi^{\rho\sigma}\raa D \partial_{\rho} U_{\sigma} 
- \laa \pi^{\mu\nu} |x_{\lambda} | I^{\rho}\raa \Delta^{\lambda\kappa}\partial_{\kappa} \partial_{\rho}(\beta \mu), \label{pi1}\\
\la \hat{I}^{\mu}\ra 
& = & \la \hat{I}^{\mu}\ra_{l} - \beta \laa I^{\mu} | I^{\nu} \raa \partial_{\nu} (\beta\mu ) 
-\laa I^{\mu}|x^{\rho}| I^{\nu} \raa \partial_{\rho} \partial_{\nu} (\beta\mu ) \nonumber \\
&&+\beta \laa I^{\mu}|x^{\nu}|\pi^{\rho\sigma} \raa  \partial_{\nu} \partial_{\rho}U_{\sigma}
-\beta \laa I^{\mu}|x^{\nu}|P'\raa  \partial_{\nu} \partial_{\rho}U_{\rho} \nonumber \\
&=& \la\hat{I}^{\mu}\ra_{l} - \beta \laa I^{\mu} | I^{\nu}\raa {\Delta_{\nu}}^{\rho}\partial_{\rho}(\beta\mu ) 
-\laa I^{\mu}|\tau| I^{\nu} \raa D \partial_{\nu}(\beta\mu ) \nonumber \\
&&+\beta \laa I^{\mu}|x^{\nu}|\pi^{\rho\sigma} \raa  {\Delta_{\nu}}^{\kappa} \partial_{\kappa} \partial_{\rho}U_{\sigma}
-\beta \laa I^{\mu}|x^{\nu}|P'\raa {\Delta_{\nu}}^{\sigma} \partial_{\sigma} \partial_{\rho}U_{\rho}, \label{I1}\\
\la \hat{P} \ra &=& \la \hat{P} \ra_{l} - \beta\laa P | P' \raa \partial_{\nu}U^{\nu} \nonumber \nonumber \\
&& -\beta \laa P | x^{\rho}|P'\raa \partial_{\rho} \partial_{\nu}U^{\nu}
-  \laa P | x^{\rho}|I^{\sigma} \raa \partial_{\rho} \partial_{\nu}(\beta \mu) \nonumber \\
&=&  \la P \ra_{l} - \beta\laa P | P' \raa \partial_{\nu}U^{\nu} \nonumber \\
&& -\beta \laa P | \tau |P'\raa D \partial_{\nu}U^{\nu}
-  \laa P | x^{\rho}|I^{\sigma} \raa {\Delta_{\rho}}^{\sigma}\partial_{\sigma} \partial_{\nu}(\beta \mu). \label{P'1}
\end{eqnarray}

For the isotropic system, we can factorize scalar coefficients from tensor structures as follows:
\begin{eqnarray}
\laa I^{i} | I^{j}\raa &=& \la I|I\ra \delta^{ij},\\
\laa \pi^{ij} | \pi^{kl} \raa &=&  \frac{\la \pi|\pi \ra }{2}\left(\delta^{ik}\delta^{jk}+\delta^{il}\delta^{jk}-\frac{2}{3}\delta^{ij}\delta^{kl}\right), \\
\laa \pi^{ij}| \tau | \pi^{kl} \raa &=& \frac{\la\pi|\bf{t}|\pi\ra}{2}\left(\delta^{ik}\delta^{jk}+\delta^{il}\delta^{jk}-\frac{2}{3}\delta^{ij}\delta^{kl}\right), \\
\laa I^{i} | \tau |I^{j}\raa &=& \la I|{\bf t}| I \ra \delta^{ij},\\
\laa P | P'\raa &=&\la P|P' \ra, \\
\laa P| \tau | P'\raa &=& \la P |{\bf t}| P' \ra, \\
\laa I^{i} | x^{j}| P' \raa &=& \la I|{\bf x}| P' \ra \delta^{ij}, \\
\laa \pi^{ij} | x^{k}| I^{l} \raa 
&=&\frac{\la\pi|{\bf x}|I \ra}{2}\left(\delta^{ik}\delta^{jk}+\delta^{il}\delta^{jk}-\frac{2}{3}\delta^{ij}\delta^{kl}\right), 
\end{eqnarray}
where scalar coefficients in the right hand side are defined by:
\begin{eqnarray}
\la\pi|\pi\ra &=& \frac{1}{5}\laa \pi^{\mu\nu}|\pi_{\mu\nu} \raa, \\
\la\pi|{\bf t}|\pi\ra &=& \frac{1}{5}\laa \pi^{\mu\nu}|\tau|\pi_{\mu\nu} \raa, \\
\la I| I\ra &=& \frac{-1}{3}\laa I^{\mu}| I_{\mu} \raa, \\
\la I|{\bf t}| I\ra &=& \frac{-1}{3}\laa I^{\mu}|\tau| I_{\mu} \raa, \\
\la\pi|{\bf x}|I \ra &=& \frac{1}{5} \laa \pi^{\mu\nu} | x_{\mu}| I_{\nu} \raa, \\
\la I|{\bf x}| P' \ra &=& \frac{-1}{3} \Delta^{\mu\nu} \laa I_{\mu} | x_{\nu}| P' \raa,\\
\la P  |{\bf x}|I \ra &=& \frac{-1}{3} \Delta^{\mu\nu} \laa P | x_{\nu}|  I_{\mu}\raa.
\end{eqnarray}
Both in equations (\ref{pi1}) and (\ref{I1}), because of the isotropy,  
the first term in the right hand side should vanish:
\begin{eqnarray}
\la \pi^{\mu\nu} \ra_{l} &=& 0,\\
\la I^{\mu} \ra_{l} &=& 0.
\end{eqnarray}
$\la \hat{P} \ra_{l}$ is static pressure. The second terms in the right hand side of 
Eqs.\ (\ref{pi1}) and (\ref{P'1}) stand for the Kubo-formulas for the shear viscosity $\eta_{s}$ and the bulk viscosity $\eta_{v}$:
\begin{eqnarray}
\eta_{s} &=& \beta \la \pi | \pi \ra, \label{kubo1a}\\
\eta_{v} &=& \beta \la P | P' \ra. \label{kubo1b}
\end{eqnarray}
The coefficient in the second term in the right hand side of Eq.\ (\ref{I1}) corresponds to the heat conductivity $\kappa$ on the E-frame\cite{LL}:
\begin{equation}
\kappa = \left( \frac{\la\hat{\varepsilon} 
\ra_{l}+\la\hat{P}\ra_{l}}{\la\hat{n}\ra_{l} T}\right)^{2}\la I|I \ra.\label{kubo1c}
\end{equation}

Finally, we can rewrite equations (\ref{pi1}),(\ref{I1}) and (\ref{P'1}) as:
\begin{eqnarray}
\la \hat{\pi}^{\mu\nu} \ra &=& \eta_{s}\mathcal{T}(^{\mu\nu}|^{\rho\sigma})\partial_{\rho}U_{\sigma}
+\beta \la\pi|{\bf t}|\pi\ra  \mathcal{T}(^{\mu\nu}|^{\rho\sigma}) D\partial_{\rho}U_{\sigma} \nonumber \\
&&-  \la\pi|{\bf x}|I \ra  \mathcal{T}(^{\mu\nu}|^{\rho\sigma}) \partial_{\rho}\partial_{\sigma}(\beta\mu),\label{pi2}
\\ \noalign{\vskip 0.4cm} 
\la \hat{I}^{\mu} \ra &=& \kappa \left( \frac{\la\hat{n}\ra_{l} T}{\la\hat{\varepsilon} 
\ra_{l}+\la\hat{P}\ra_{l}}
\right)^2 \Delta^{\mu\nu}\partial_{\nu}(\beta\mu)+
 \la I|{\bf t}| I \ra \Delta^{\mu\nu}D\partial_{\nu}(\beta\mu)  \nonumber \\ \noalign{\vskip 0.4cm} 
&&+
\beta \la I|{\bf x}| \pi \ra \mathcal{T}(^{\mu\nu}|^{\rho\sigma}) \partial_{\nu}\partial_{\rho}U_{\sigma},\label{I2} 
+ \beta\la I|{\bf x}| P' \ra \Delta^{\mu\nu}\partial_{\nu} \partial_{\sigma}U^{\sigma}, \label{I2} 
 \\ \noalign{\vskip 0.4cm} 
\la \hat{P} \ra - \la \hat{P} \ra_{l} &=&
-\eta_{v}\partial_{\mu}U^{\mu} - \beta \la P |{\bf t}|P' \ra D \partial_{\mu}U^{\mu} 
+  \la P |{\bf x}| I \ra \Delta^{\mu\nu}\partial_{\mu}\partial_{\nu}(\beta \mu).\label{P'2}
\end{eqnarray}
Substituting the first order equation (the Navier-Stokes equation):
\begin{eqnarray}
\la \hat{\pi}^{\mu\nu} \ra &=& \eta_{s}\mathcal{T}(^{\mu\nu}|^{\rho\sigma})\partial_{\rho}U_{\sigma}, \nonumber \\ \noalign{\vskip 0.4cm} 
\la \hat{I}^{\mu} \ra &=& \kappa \left( \frac{\la\hat{n}\ra_{l} T}{\la\hat{\varepsilon} 
\ra_{l}+\la\hat{P}\ra_{l}}
\right)^2 \Delta^{\mu\nu}\partial_{\nu}(\beta\mu),  
\nonumber \\ \noalign{\vskip 0.4cm} 
\la \hat{P} \ra - \la \hat{P} \ra_{l} &=&
-\eta_{v}\partial_{\mu}U^{\mu}, \nonumber 
\end{eqnarray}
for the second order derivative terms of $U^{\mu}$ and $\beta\mu$, 
we can obtain the Israel-Stewart equation:
\begin{eqnarray}
\la \hat{\pi}^{\mu\nu} \ra &=& \eta_{s}\mathcal{T}(^{\mu\nu}|^{\rho\sigma})\partial_{\rho}U_{\sigma}
+ \frac{\beta \la\pi|{\bf t}|\pi\ra}{\eta_{s} } 
\mathcal{T}(^{\mu\nu}|^{\rho\sigma}) D\la \hat{\pi}^{\rho\sigma} \ra 
\nonumber \\ &&
- \frac{ \la\pi|{\bf x}|I \ra }
{ \kappa \left( \frac{\la\hat{n}\ra_{l} T}{\la\hat{\varepsilon}\ra_{l} + \la\hat{P}\ra_{l}}
\right)^2 }\mathcal{T}(^{\mu\nu}|^{\rho\sigma}) \partial_{\rho}\la \hat{I}_{\sigma} \ra,
\label{pi3} \\ \noalign{\vskip 0.4cm} 
\la \hat{I}^{\mu} \ra &=& \kappa \left( \frac{\la\hat{n}\ra_{l} T}{\la\hat{\varepsilon} \ra_{l}+\la\hat{P}\ra_{l}}
\right)^2 \Delta^{\mu\nu}\partial_{\nu}(\beta\mu) 
+  \frac{\la I|{\bf t}| I \ra}{ \kappa \left( \frac{\la\hat{n}\ra_{l} T}{\la\hat{\varepsilon}\ra_{l}+\la\hat{P}\ra_{l}}
\right)^2 } \Delta^{\mu\nu}D \la \hat{I}_{\nu}  \ra
 \nonumber \\ \noalign{\vskip 0.4cm} 
&&+\frac{\beta \la I|{\bf x}| \pi \ra} { \eta_{s} }
\mathcal{T}(^{\mu\nu}|^{\rho\sigma}) \partial_{\nu}\la \hat{\pi}_{\rho\sigma} \ra
+ \frac{\beta \la I|{\bf x}| P' \ra }{-\eta_{v}}\Delta^{\mu\nu}\partial_{\nu}\left(\la \hat{P} \ra -\la \hat{P} \ra_{l}\right) , 
\label{I3} \\
\noalign{\vskip 0.4cm} 
\la \hat{P} \ra - \la \hat{P} \ra_{l} &=&
-\eta_{v}\partial_{\mu}U^{\mu} + \frac{\beta \la P |{\bf t}|P' \ra}{\eta_{v} } D 
\left(\la \hat{P} \ra - \la \hat{P} \ra_{l} \right)  
+  \frac{\la P |{\bf x}| I \ra }{\kappa \left( \frac{\la\hat{n}\ra_{l} T}{\la\hat{\varepsilon}\ra_{l}+\la\hat{P}\ra_{l}}
\right)^2 }
\Delta^{\mu\nu}\partial_{\mu}\la \hat{I}_{\nu} \ra. \label{P'3}
\end{eqnarray}

Comparing the above  equations (\ref{pi3}), (\ref{I3}) and (\ref{P'3}) with  (2.38) in ref \cite{IS2}, 
we can identify the $\alpha_i$'s and $\beta_i$'s in the Israel-Stewart equation in our notation.  
Besides thermodynamical quantities and the numerical factors coming from the differences in
the definition, $\beta_{i}$'s in ref. \cite{IS2} are current-weighted correlation times; $\beta_{0} \sim \la P|{\bf t}|P'\ra$, $\beta_{1} \sim \la I|{\bf t}|I \ra$ and $\beta_{2} \sim \la\pi|{\bf t}|\pi\ra$.  
 $\alpha_{i}$ in ref. \cite{IS2} is cross-current-weighted correlation length; $\alpha_{1} \sim \la I|{\bf x}|\pi \ra $. As concerns bulk current, though the original current operator is $\hat{P}$ the thermodynamical current in $\hat{B}$ is $\hat{P'}$, hence, $\alpha_{0} \sim \la P|{\bf x}| I \ra $ in bulk current (2.38a) and $-\alpha_{0} \sim \la I|{\bf x}| P' \ra $ in charge current (2.38b) in ref. \cite{IS2}.

Normalizing the "state" $|\pi\ra$,  $|I \ra$ and $|P' \ra$ by using the relation,
  (\ref{kubo1a}),  (\ref{kubo1b}) and  (\ref{kubo1c}), 
we define the current-weighted correlation times:  
 \begin{eqnarray}
 \bar{\tau}_{I} &=& \frac{\la I|{\bf t }|I \ra}{\la I|I \ra}, \\
 \bar{\tau}_{s} &=& \frac{\la \pi|{\bf t }|\pi \ra}{\la \pi|\pi \ra}, \\
 \bar{\tau}_{v} &=& \frac{\la P|{\bf t }|P' \ra}{\la P|P' \ra}, 
\end{eqnarray}
and the cross-current weighted correlation distances:
\begin{eqnarray}
\bar{\bf x}_{Is} &=& 
\left(\frac{\la\hat{\varepsilon}\ra_{l}+\la\hat{P}\ra_{l}}{\la\hat{n}\ra_{l} T} \right) 
\sqrt{\frac{1}{\kappa}} \sqrt{\frac{\beta}{\eta_{s}}}
\la I|{\bf x }|\pi \ra, \\
\bar{\bf x}_{Iv} &=& 
\left(\frac{\la\hat{\varepsilon}\ra_{l}+\la\hat{P}\ra_{l}}{\la\hat{n}\ra_{l} T} \right) 
\sqrt{\frac{1}{\kappa}} \sqrt{\frac{\beta}{\eta_{v}}}
\la I|{\bf x }| P'\ra, \\
 \bar{\bf x}_{vI} &=& 
\left(\frac{\la\hat{\varepsilon}\ra_{l}+\la\hat{P}\ra_{l}}{\la\hat{n}\ra_{l} T} \right) 
\sqrt{\frac{1}{\kappa}} \sqrt{\frac{\beta}{\eta_{v}}}
\la P|{\bf x }| I\ra. 
\end{eqnarray}
All these quantities are the functions of the temperature $T$ and the chemical potential $\mu$, 
which are able to be calculated based on the statistical 
mechanics in equilibrium.

\section{Concluding Remarks}

Based on the nonequilibrium density operator, we have derived the hyperbolic hydrodynamical equation. 
 All coefficients in the equations are expressed as
the integration of the canonical correlations of the current operators which are to be 
calculated in statistical mechanics.  Our discussion is nothing special 
for the relativity but we simply expand the thermodynamical forces in their derivatives. 
The additional coefficients in the Israel-Stewart equation $\alpha_i$'s and $\beta_i$'s are  given as the current-weighted correlation times and 
the cross-current-weighted correlation 
distances.

The current-weighted correlation time $\beta_i$ corresponds to the relaxation time of the current.
The evaluation of the relaxation time 
$\beta_i$ is essentially the same to the calculation of the Kubo-formulas for 
the transport coefficient\cite{Xian}.  
According to the hadro-molecular simulation, each current exhibits it's own relaxation even if the basic dynamics is common and once we succeed to figure out 
the behavior of the relaxation of the currents, we can easily evaluate both the relaxation time $\beta_i$ and the transport coefficient\cite{SMMN}.  This method has also been applied to the calculation of the viscosity of a hadron gas\cite{UrQMD, MS}.
$\alpha_i$'s are the correlation distances between different currents 
which have not yet been investigated.  

As Iso et.al. discussed, the comparison between the microscopic correlation length and macroscopic scale in hydrodynamics is the touch-stone of the model \cite{IMN}.
The canonical correlations exhibit typical microscopic scales of the system.
The evaluation of  $\alpha_i$ and $\beta_i$ or quantities in Eqs.\ (62)$\sim$(67) based on the Lattice QCD, or hadro-molecular calculation, will provide us the key informations to justify the hydrodynamic model of the QCD matter. We have treated only the linear term of 
the thermodymical forces, but higher order purturbation or non-perturbative 
treatment would be
appreciated in the microscopic calculation of the canonical correlation.

The bulk current, Eq.\ (\ref{P'}) is a natural extension of the $\hat{P}'$ in \cite{CH} and \cite{Zuba}
 where the charge conservation is neglected.  
If a conserved current exists, $\hat{P} \propto  \hat{\varepsilon}$ is not enough to make  $\hat{P}'$ vanish.  
Therefore, even if $ \la \hat{P} \ra = \frac{1}{3} \la \hat{\varepsilon}\ra $ is achieved in a high energy region, the vanishing of the bulk viscosity may not be trivial.

\section*{Acknowledgments}
The authors would like to thank Prof.\ Tetsufumi Hirano and Prof.\ Naomichi Suzuki for fruitful discussion.  One of the authors (M.~M.~) acknowledges  the encouragement of Prof. Hiromichi Nakazato and Prof. Hiroyuki Abe.
This work is supported by Grants-in-Aid for Research Activity of Matsumoto University No.\ 12111048.

\bibliography{apssamp}

\end{document}